\documentclass[floats,prd,aps,amssymb,nofootinbib]{revtex4}
\usepackage{amssymb}
\usepackage{graphics}
\usepackage{graphicx}
\usepackage{amsmath}
\usepackage{color}
\usepackage{amsfonts}
\usepackage{bm}
\usepackage[]{latexsym}
\usepackage{float}

\newcommand{\be}{\begin{equation}}\newcommand{\ee}{\end{equation}}
\newcommand{\bea}{\begin{eqnarray}}\newcommand{\eea}{\end{eqnarray}}
\newcommand{\brr}{\begin{array}}\newcommand{\err}{\end{array}}
\newcommand{\bit}{\begin{itemize}}\newcommand{\eit}{\end{itemize}}
\newcommand{\ben}{\begin{enumerate}}\newcommand{\een}{\end{enumerate}}

\newcommand{\ba}{\begin{array}}
\newcommand{\ea}{\end{array}}

\def\lf{\left}

\def\ri{\right}
\def\al{\alpha}

\def\1{{_{1}}}\def\2{{_{2}}}

\begin{document}
\title{Spontaneous Lorentz violation from infrared gravity}

\author{Fabrizio Illuminati\footnote{filluminati@unisa.it}$^{\hspace{0.3mm}1,2}$, Gaetano Lambiase\footnote{lambiase@sa.infn.it}$^{\hspace{0.3mm}2,3}$ and Luciano Petruzziello\footnote{lupetruzziello@unisa.it}$^{\hspace{0.3mm}1,2}$} \affiliation
{$^1$Dipartimento di Ingegneria Industriale, Universit\`a degli Studi di Salerno, Via Giovanni Paolo II, 132 I-84084 Fisciano (SA), Italy.\\ $^2$INFN, Sezione di Napoli, Gruppo collegato di Salerno, Italy. 
\\ $^3$Dipartimento di Fisica, Universit\`a degli Studi di Salerno, Via Giovanni Paolo II, 132 I-84084 Fisciano (SA), Italy.}

\date{\today}
\def\be{\begin{equation}}
\def\ee{\end{equation}}
\def\al{\alpha}
\def\bea{\begin{eqnarray}}
\def\eea{\end{eqnarray}}

\begin{abstract}
In this paper, we investigate a novel implication of the non-negligible spacetime curvature at large distances when its effects are expressed in terms of a suitably modified form of the Heisenberg uncertainty relations. Specifically, we establish a one-to-one correspondence between such modified uncertainty principle and the Standard Model Extension (SME), a string-theoretical effective field theory that accounts for {both explicit and spontaneous} breaking of Lorentz symmetry. This tight correspondence between string-derived effective field theory and modified quantum mechanics with extended uncertainty relations is validated by comparing the predictions concerning a deformed Hawking temperature derived from the two models. Moreover, starting from the experimental bounds on the gravity sector of the SME, we derive the most stringent constraint achieved so far on the value of the free parameter entering in the extended Heisenberg uncertainty principle.
\end{abstract}

\vskip -1.0 truecm 

\maketitle

\section{Introduction}

Since the establishment of quantum mechanics and general relativity, there has been a constantly growing effort to merge quantum and gravitational effects at arbitrary energy scales in a complete and consistent theoretical framework. This effort has produced several plausible candidates for a quantum theory of the gravitational interaction such as string theory~\cite{string}, loop quantum gravity~\cite{lqg,lqg2}, and doubly special relativity~\cite{dsr,dsr2,dsr3}. The study of the interplay between quantum and gravitational effects appears to be particularly important at the currently reachable energies, as this is the regime where we have hope that physical predictions can factually be probed in tabletop laboratory tests. Therefore, the analysis of low-energy, infrared (IR) quantum gravitational manifestations may represent a promising starting point for the possible construction of ultraviolet (UV) complete models of quantum gravity. In this respect, an additional motivation for the study of IR phenomena stems from the UV/IR duality discovered for the first time in the context of the AdS/CFT correspondence~\cite{adscft,adscft2,adscft3}.

In the present paper, we focus on IR gravity to exhibit how Lorentz symmetry is affected by the non-negligible spacetime curvature at large distances. In a quantum mechanical setting, such a feature can be incorporated by extending the Heisenberg uncertainty principle (HUP) with the addition of a position-dependent correction that introduces a non-vanishing minimal uncertainty in momentum~\cite{kempf,kempf2,cbl,oy,mz} and provides a form of an extended uncertainty principle (EUP). Among different versions of the EUP, the best known one includes a universal modification that is geometric-independent; indeed, such a generalization of the standard uncertainty relations arises naturally when merging quantum mechanics and general relativity. In other scenarios, the version of the EUP that is accounted for entails a dependence on the intrinsic geometric properties of the underlying background curvature; for more details along this direction, see Refs.~\cite{schur,schur2,fab,fab2,fab3}.
For the sake of completeness, it is worth mentioning that several works inspired by string theory~\cite{Amati87,Kempf95,Maggiore93,Adler01,AdlerDuality} also allow for the existence of a momentum-dependent correction to the HUP; such an extension is known as the generalized uncertainty principle (GUP) (for several development of this subject, see Refs.~\cite{noncom2,noncom3,bis,quar,ses,ong,set,ot,dieci,qft,qft4,plenio2,dasprl,qgd} and therein).

In order to quantify the breakdown of Lorentz symmetry induced by the EUP, we consider a string-theoretical effective field theory according to which any operator appearing in the Standard Model (SM) Lagrangian is contracted with Lorentz-violating fields~\cite{sme1}. This model is known as the Standard Model Extension (SME), and we will focus in particular on its gravity sector~\cite{sme2} in the non-relativistic limit~\cite{sme3}.
Specifically, along the line of Ref.~\cite{ls}, in the following we establish a one-to-one correspondence between the predictions of the EUP and of the SME concerning the deformation of the Hawking temperature for a Schwarzschild black hole
\be\label{ht}
T_H=\frac{\hbar c^3}{8\pi k_B G M}\,.
\ee
In so doing, we essentially relate the free deformation parameter arising in the framework of the EUP with the Lorentz-violating fields contained in the SME Lagrangian, thereby explaining spontaneous Lorentz symmetry breaking in terms of large scale effects. Moreover, by relying on the experimental bounds associated with the SME gravity sector, we manage to derive novel constraints on the EUP free deformation parameter which are more stringent than the previously known ones~\cite{mz}.

The paper is organized as follows: in Sec.~II, we briefly give an overlook of the main aspects of EUP together with a heuristic derivation of the modified Hawking temperature. The same procedure will be carried out for the SME in Sec.~III, whereas Sec.~IV contains the theoretical and numerical comparison between the two predictions. Finally, Sec.~V contains concluding remarks and discussions.

\section{Extended uncertainty principle and modified Hawking temperature}

Starting from the Heisenberg uncertainty principle
\be\label{hup}
\Delta x\Delta p\geq\frac{\hbar}{2}\,,
\ee
one can incorporate the influence of spacetime curvature at large distances by adding a position-dependent term in the r.h.s. of Eq.~\eqref{hup}, namely~\cite{kempf,cbl,oy,mz}
\be\label{eup}
\Delta x\Delta p\geq\frac{\hbar}{2}\lf(1+\al\Delta x^2\ri)\,,
\ee
with $\al$ being the inverse of a squared length and $\al\Delta x^2\ll1$. One possible interpretation of the free deformation parameter $\al$ is to conceive it as a function of the cosmological constant $\Lambda$ in a (anti-) de Sitter space~\cite{mignemi}. However, in greater generality it can simply be seen as a consequence of the intrinsic spacetime curvature at large cosmological distances that enforces a limit to the precision with which to resolve the momentum of a point particle~\cite{mz}. In turn, this fact implies that there exists a minimal uncertainty for $p$ proportional to the constant $\al$, i.e. $\Delta p_{min}\simeq\hbar\sqrt{|\al|}$. {Note that the parameter $\al$ does not have to be necessarily constant; indeed, it may be a function of spacetime position or even a stochastic variable. Similar considerations have already been addressed in the context of GUP~\cite{ong,qgd} and they equally hold true for the EUP currently investigated.}

The above inequality~\eqref{eup} can be straightforwardly derived from the deformed canonical commutation relation
\be\label{dcr}
\lf[\hat{X},\hat{P}\ri]=i\hbar\lf(\textbf{1}+\al\hat{X}^2\ri)\,.
\ee
From the previous equation, one can deduce a simple representation of $\hat{X}$ and $\hat{P}$ in terms of auxiliary operators $\hat{x}$ and $\hat{p}$ for which the standard canonical commutation relation holds (i.e. $\lf[\hat{x},\hat{p}\ri]=i\hbar$). Specifically:
\be\label{rep}
\hat{X}=\hat{x}\,, \qquad\qquad \hat{P}=\lf(1+\al\hat{x}^2\ri)\hat{p}-2i\al\hbar\hat{x}\,,
\ee
where the second term in the r.h.s. of the second expression ensures that $\hat{P}=\hat{P}^\dagger$. {The three-dimensional analysis of the above framework would allow for the emergence of spatial non-commutativity~\cite{kempf,kempf2}, since one can immediately verify that $[\hat{X}_j,\hat{X}_k]\neq0$. Nevertheless, we can safely ignore all the non-commutative corrections associated with the operators $\hat{X}_j$, as they would depend on higher powers of $\al$, i.e.
\be
\hat{X}_j=\hat{x}_j+\mathcal{O}(\al^2)\,,
\ee
 which in the present case are neglected.} We refer to Refs.~\cite{kempf,kempf2} for further mathematical details.

To evaluate the deformed Hawking temperature of a Schwarzschild black hole of mass $M$, we follow some simple heuristic considerations as outlined, e.g., in Refs.~\cite{bis,heuristic}. The starting point is the natural assumption that the position uncertainty of the photons just after they have been emitted by the black hole is proportional to the Schwarzschild radius, namely $\Delta x\simeq\gamma r_s$, with $\gamma$ to be fixed by requiring consistency with the standard picture in the limit $\al\to0$. Under these circumstances, from Eq.~\eqref{eup} we obtain 
\be\label{step1}
\Delta p\simeq\frac{\hbar c^2}{4\gamma G M}\lf[1+4\al\frac{G^2M^2}{c^4}\gamma^2\ri]\,.
\ee
Now, we can express the characteristic energy of the emitted photons $\Delta pc$ in terms of the temperature of the radiation in compliance with the equipartition theorem~\cite{bis,heuristic}, by virtue of which $\Delta pc\simeq k_BT$. Finally, the equation for the $\al$-deformed Hawking temperature $T_{EUP}$ reads
\be\label{heup}
T_{EUP} = \,\frac{\hbar c^3}{4\gamma k_B G M}\lf[1+4\al\frac{G^2M^2}{c^4}\gamma^2\ri],
\ee
which, in terms of the Hawking temperature $T_{H}$ reads:
\be\label{heup2}
T_{EUP} =T_{H}\lf[1+16\al\pi^2\frac{G^2M^2}{c^4}\ri],
\ee
where we have set $\gamma=2\pi$ in order to recover the original Hawking result in the limit $\al\to0$. For a thorough discussion on the above correspondence, we refer the reader to Ref.~\cite{hc}.

\section{Standard Model Extension and modified Hawking temperature}

The Standard Model Extension is a generalization of the Standard Model of particle physics which predicts {both explicit and spontaneous} Lorentz symmetry breaking. Motivated by string-theoretical arguments~\cite{sme1}, the SME enlarges the SM domain by contracting any SM field with Lorentz-violating operators which give rise to new  phenomenology. {Although the Standard Model Extension was originally conceived to extend the Standard Model only, later on also the gravitational interaction has been added with the ensuing Lorentz-violating coefficients.}

For our purposes, we are interested in investigating the {minimal} SME gravity sector~\cite{sme2,sme3,ls}, {which includes exclusively Lorentz-violating operators of mass dimension three or four.} In particular, by denoting with $S_{EH}$ and $S_m$ the Einstein-Hilbert and the matter action respectively, the {minimal} SME total gravitational action reads~\cite{sme3}
\be\label{action}
S=S_{EH}+S_m+S_{LV}\,, \qquad S_{LV}=\frac{c^4}{16\pi G}\int d^4x\sqrt{-g}\lf(-uR+s^{\mu\nu}R^T_{\mu\nu}+t^{\mu\nu\rho\lambda}C_{\mu\nu\rho\lambda}\ri)\,,
\ee
where $S_{LV}$ denotes the effective Lorentz symmetry breaking action derived from string theory, $R$ is the Ricci scalar, $R^T_{\mu\nu}$ is the traceless Ricci tensor, $C_{\mu\nu\rho\lambda}$ is the Weyl conformal tensor, and $u$, $s^{\mu\nu}$ and $t^{\mu\nu\rho\lambda}$ are the Lorentz-violating effective fields. 

{In the regime of the} post-Newtonian (PPN) approximation~\cite{ppn}, {a Schwarzschild-like solution of the linearized field equations for the minimal SME can be found~\cite{sme3}, and it is given by}
\be\label{linel}
ds^2=f(r)c^2dt^2-\frac{1}{f(r)}dr^2-r^2d\Omega^2 \, ,
\ee
where
\be\label{linel2}
f(r)=1-\frac{2GM}{rc^2}\lf[1+\bar{s}^{ij}g_{ij}(\theta,\phi)\ri]\,.
\ee
In the above, $\bar{s}^{ij}$ denote the vacuum expectation values of the fields $s^{ij}$ and $g_{ij}(\theta,\phi)$ are functions of the angular coordinates, for which the inequality $g_{ij}(\theta,\phi)\leq1$ holds regardless of the choice for $\theta$ and $\phi$. {The absence of $\bar{u}$ in Eq.~\eqref{linel2} is related to the fact that a non-vanishing value for such parameter only amounts to a scaling of the PPN metric~\cite{sme3}, and thus it can be set to zero. On the other hand, the disappearance of $\bar{t}^{\mu\nu\rho\lambda}$ is a typical feature occurring in the post-Newtonian SME gravitational phenomenology known as ``t puzzle'', and it has been extensively discussed in several works (see for instance Refs.~\cite{sme4,sme5,sme6,sme7,sme8,sme9,sme10}).}

In order to evaluate the deformed Hawking temperature $T_{SME}$ arising in the SME framework, we must compute~\cite{ls,bh}
\be\label{hsme}
T_{SME} = \frac{\hbar c}{4\pi k_B}\frac{df(r)}{dr}\Bigr|_{r=r_0}\,,
\ee
where $r_0$ solves the equation $f(r_0)=0$, thereby denoting the horizon radius. At this point, a straightforward calculation yields
\be\label{hsme2}
T_{SME}=T_H\Bigl[1-\bar{s}^{ij}g_{ij}(\theta,\phi)\Bigr]\,,
\ee
which has to be compared with Eq.~\eqref{heup2} derived in the EUP framework. {Before concluding this Section, it is worth pointing out that the temperature~\eqref{hsme2} is anisotropic, as it explicitly depends on the angular position. However, such an observation does not undermine the validity of our main goal; indeed, Eq.~\eqref{heup2} might be anisotropic as well, since $\al$ is not bound to be a constant.}

\section{Comparison and consistency conditions}

We will now look at the relations that are required in order to achieve consistency between the predictions~\eqref{heup2} and \eqref{hsme2} and thus relate the large scale effects of spacetime curvature with spontaneous breaking of the Lorentz symmetry. The comparison of the deformed Hawking temperatures deduced from the two distinct physical settings shows that the two are consistent provided that the following identification holds:
\be\label{res}
\al=-\frac{c^4}{16\pi^2G^2M^2}\bar{s}^{ij}g_{ij}(\theta,\phi)\,,
\ee
for fixed values of $\theta$ and $\phi$. {Therefore, as already argued, the magnitude of $\al$ is not constant, but it varies with the angular displacement, thus giving rise to an anisotropic $T$ also for the EUP-corrected Hawking temperature.} Since the $g_{ij}(\theta,\phi)$ can always be taken as positive quantities~\cite{ls}, the sign of $\al$ strictly depends on the sign of the Lorenz-violating coefficients $\bar{s}^{ij}$. Additionally, {by means of consistency arguments that do not require Lorentz invariance}, it is known that $\al$ correctly characterizes an EUP associated with an expanding universe if and only if $\al < 0$~\cite{cbl,oy,mignemi}. Therefore, to allow agreement with experimental evidences, assuming $g_{ij}(\theta,\phi)\simeq1\,$ $\forall i,j$~\cite{ls}, we must necessarily impose 
\be\sum_{i,j}\bar{s}^{ij}>0,\ee 
which realizes a novel bound on the admissible values of the Lorentz-violating coefficients. Of course, by virtue of Eq.~\eqref{res}, the bounds holding for the quantities $\bar{s}^{ij}$ in turn imply bounds on the EUP free deformation parameter that lead to an extremely significant improvement with respect to the existing constraints~\cite{mz} on the possible values of $\al$. 

In Table 1 we report the known bounds on $\bar{s}^{ij}$, the ensuing constraints they entail on $\sqrt{|\al|}$ by virtue of Eq.~\eqref{res} and the corresponding experimental frameworks used to determine each bound.

\begin{table}[ht]
\caption{Estimated bounds on the EUP parameter} % title of Table
\centering % used for centering table
\begin{tabular}{c c c} % centered columns (4 columns)
\hline\hline %inserts double horizontal lines
Experiments &\hspace{4mm} Bounds on $|\bar{s}^{ij}|$ &\hspace{4mm} Bounds on $\sqrt{|\al|}$ \\ [0.5ex] % inserts table
%heading
\hline \vspace{1mm}
Geodetic effect ($M=M_\oplus$)~\cite{rev} &\hspace{2.5mm} $|\bar{s}^{ij}|\lesssim10^{-3}$ &\hspace{4mm} $\sqrt{|\al|}\lesssim6.37\times10^{-2}\,m^{-1}$ \\ \vspace{1mm}
Gravity Probe B ($M=M_\oplus$)~\cite{rev} &\hspace{2.5mm} $|\bar{s}^{ij}|\lesssim10^{-4}$ &\hspace{4mm} $\sqrt{|\al|}\lesssim2.01\times10^{-2}\,m^{-1}$ \\ \vspace{1mm}
Frame dragging ($M=M_\oplus$)~\cite{rev} &\hspace{2.6mm} $|\bar{s}^{ij}|\lesssim10^{-7}$ &\hspace{4mm} $\sqrt{|\al|}\lesssim6.37\times10^{-4}\,m^{-1}$ \\ \vspace{1mm}
Gravimetry ($M=M_\oplus$)~\cite{rev} &\hspace{4.4mm} $|\bar{s}^{ij}|\lesssim10^{-10}$ &\hspace{4mm} $\sqrt{|\al|}\lesssim2.01\times10^{-5}\,m^{-1}$ \\ \vspace{1mm}
Lunar laser ranging ($M=M_\oplus$)~\cite{rev,llr} &\hspace{4.3mm} $|\bar{s}^{ij}|\lesssim10^{-12}$ &\hspace{4mm} $\sqrt{|\al|}\lesssim2.01\times10^{-6}\,m^{-1}$ \\ \vspace{1mm}
Torsion pendulum ($M=M_\oplus$)~\cite{rev} &\hspace{4.4mm} $|\bar{s}^{ij}|\lesssim10^{-15}$ &\hspace{4mm} $\sqrt{|\al|}\lesssim6.37\times10^{-8}\,m^{-1}$ \\ \vspace{1mm}
Perihelion precession ($M=M_\odot$)~\cite{rev} &\hspace{3mm} $|\bar{s}^{ij}|\lesssim10^{-9}$ &\hspace{5.7mm} $\sqrt{|\al|}\lesssim1.98\times10^{-10}\,m^{-1}$ \\ \vspace{1mm}
Binary pulsar ($M=2.8\,M_\odot$)~\cite{rev,bp} &\hspace{4.4mm} $|\bar{s}^{ij}|\lesssim10^{-11}$ &\hspace{6mm} $\sqrt{|\al|}\lesssim7.05\times10^{-12}\,m^{-1}$ \\ \vspace{1mm}
Solar-spin precession ($M=M_\odot$)~\cite{rev} &\hspace{4.5mm} $|\bar{s}^{ij}|\lesssim10^{-13}$ &\hspace{5.9mm} $\sqrt{|\al|}\lesssim1.98\times10^{-12}\,m^{-1}$ \\ [1ex] % [1ex] adds vertical space
\hline\hline
\end{tabular}
\label{table} % is used to refer this table in the text
\end{table}

The bounds on $\sqrt{|\al|}$ derived from the geodetic effect and Gravity Probe B are similar to the ones typically encountered in phenomenological works on this topic~\cite{mz}. Consequently, we note that all the other results contained in Table 1 considerably strengthen the constraint on $\al$. Specifically, the inequality $\sqrt{|\al|}\lesssim1.98\times10^{-12}\,m^{-1}$ provides the best current bound on the EUP free deformation parameter $\al$, and further refinement of the experimental sensitivity may allow for an even more stringent constraint. 

It is worth observing that, should one regard $\al$ as emergent from a non-vanishing cosmological constant in de Sitter space, we would have $\al=-\Lambda/3\simeq-3.66\times10^{-53}\,m^{-2}$, which in the framework of a near-Earth experiment would correspond to $|\bar{s}^{ij}|\simeq9.06\times10^{-54}$. This is in line with the expectation that Lorentz-violating coefficients should be indeed extremely small corrections to Lorentz-symmetric physics~\cite{sme3}.

\section{Concluding remarks}

In this work, we have investigated the consequences of relating large scale effects ascribable to the non-negligible spacetime curvature and the spontaneous Lorentz symmetry breaking as described by the gravity sector of the Standard Model Extension. The relation is obtained by requiring consistency between the different modifications of the Hawking temperature predicted by the SME and by a quantum mechanical model endowed with an extended uncertainty principle deformed due to spacetime curvature effects. Investigating the consequences of the consistency relations imposed between the SME and the EUP, we have shown how to significantly enhance the existing bounds on the EUP curvature-induced deformation parameter starting from the experimental constraints on the Lorentz-violating coefficients that enter the gravity sector of the SME. This simple comparison points at the possibility, in suitable settings, of probing high-energy quantum physics via low-energy gravitational effects. To some extent, the idea underlying the present study shares the same philosophy characterizing the well-known AdS/CFT correspondence~\cite{adscft}, as it may potentially provide some further insight towards a full understanding of the UV/IR duality.

\end{document}